\documentclass{svjour3}                     % onecolumn (standard format)
\smartqed  % flush right qed marks, e.g. at end of proof
\usepackage{graphicx}
\begin{document}

\title{Measuring Black Hole Spin using X-ray Reflection Spectroscopy %\thanks{Grants or other notes
}
%\subtitle{A pedagogical review}


\author{Christopher S. Reynolds
}


\institute{C.S. Reynolds \at
              Dept. of Astronomy,University of Maryland,College Park, MD20742, USA \\
              Tel.: +1-301-405-2682; 
              Fax: +1-301-314-9067; 
              \email{chris@astro.umd.edu}           %  \\
}

\date{Received: date / Accepted: date}

\maketitle

\begin{abstract}
I review the current status of X-ray reflection (a.k.a. broad iron line) based black hole spin measurements.  This is a powerful technique that allows us to measure robust black hole spins across the mass range, from the stellar-mass black holes in X-ray binaries to the supermassive black holes in active galactic nuclei. After describing the basic assumptions of this approach, I lay out the detailed methodology focusing on ``best practices" that have been found necessary to obtain robust results.  Reflecting my own biases, this review is slanted towards a discussion of supermassive black hole (SMBH) spin in active galactic nuclei (AGN).  Pulling together all of the available {\it XMM-Newton} and {\it Suzaku} results from the literature that satisfy objective quality control criteria, it is clear that a large fraction of SMBHs are rapidly-spinning, although there are tentative hints of a more slowly spinning population at high ($M>5\times 10^7M_\odot$) and low ($M<2\times 10^6M_\odot$) mass.  I also engage in a brief review of the spins of stellar-mass black holes in X-ray binaries.  In general, reflection-based and continuum-fitting based spin measures are in agreement, although there remain two objects (GRO~J1655--40 and 4U~1543--475) for which that is not true.  I end this review by discussing the exciting frontier of relativistic reverberation, particularly the discovery of broad iron line reverberation in {\it XMM-Newton} data for the Seyfert galaxies NGC~4151, NGC~7314 and MCG--5-23-16.  As well as confirming the basic paradigm of relativistic disk reflection, this detection of reverberation demonstrates that future large-area X-ray observatories such as {\it LOFT} will make tremendous progress in studies of strong gravity using relativistic reverberation in AGN.
\end{abstract}

\section{Introduction}
\label{intro}

Despite their exotic nature, General Relativity (GR) tells us that black holes are remarkably simple objects, at least as far as the external Universe is concerned. In astrophysical settings (where charge neutrality is essentially guaranteed and hence the spacetime is described by the Kerr metric; Kerr 1963), an isolated black hole is fully characterized by just its mass ($M$) and angular momentum (usually specified in terms of the dimensionless spin parameter $a=Jc/GM^2$, where $J$ is the angular momentum and the cosmic censorship hypothesis ensures that $|a|\le 1$). For gravitational processes, the mass acts as a trivial scaling factor for distances, timescales, and energies.  This directly underlies the fact that accreting black holes across the entire mass scale (from those in X-ray binaries [BH-XRBs] to active galactic nuclei [AGN]) have very similar spectral and temporal characteristics, once one scales the energies and timescales appropriately.

Spin, however, is a much more interesting quantity.  Depending upon the spin, the Kerr metric can possess qualitatively different properties that have astrophysical significance.  For example, for slowly rotating black holes, the innermost stable circular orbit (ISCO) is far from the ergosphere; accreting matter will be well into its ballistic ``death plunge" before encountering the ergosphere and the strong frame dragging effects.  But for $a>0.93$, the ISCO moves into the ergosphere raising the possibility that the accretion disk can tap into and radiate the rotational energy of the black hole (Agol \& Krolik 2000).  Indeed, the current paradigm is that relativistic jets from black holes are powered by the magnetic extraction of rotational energy from a central Kerr black hole (Blandford \& Znajek 1977). 

Clearly, measuring black hole spin is of great interest.  It is a necessary step in any Kerr-metric based tests of strong-field GR.   Spin measurements are also vital if we are to understand the astrophysical consequences of spin and, particularly, give observational grounding to theoretical notions such as the spin-driving of relativistic jets.    Measuring spin is a challenging endeavour, however --- to find the signatures of spin, we must search for and characterize (with high precision) signals from within a few gravitational radii of the black hole.

In this Chapter, I review the progress that has been made in measuring black hole spins using relativistic X-ray reflection spectroscopy (a.k.a. broad iron line spectroscopy).  With this one technique, we can study black hole spin across the mass range, from the stellar-mass black holes in X-ray binaries to the supermassive black holes in active galactic nuclei (AGN).   A major advantage of this method is that we require no knowledge of the black hole mass or its distance in order to derive the black hole spin.  Neither do we require knowledge of the accretion disk inclination --- indeed, X-ray reflection spectroscopy allows us to measure the inclination concurrently with the black hole spin.     

In contrast to the continuum fitting (CF) method for measuring spins (McClintock \& Narayan, this volume), the relativistic reflection method has been developed, evolved and debated by several groups over the past few years, producing a literature that can be rather bewildering.  Hence, in this Chapter, I review the assumptions on which the technique rests, summarize the current ``best practices" for employing this technique, and describe how careful consideration of the astrophysical consistency (and, in some cases, instrumental issues compromising the data) can resolve discrepancies in published spins.  Reflecting my own biases, much of this discussion will be dedicated to supermassive black holes (SMBHs), but I shall also describe recent work on stellar-mass black holes in BH-XRBs and the address the concordance between relativistic reflection and CF spin measurements.  I shall end by discussing recent results on relativistic reverberation.  

This Chapter is organized as follows.  Section~\ref{sec:methodology} lays out the assumptions and methodology of the relativistic reflection spin measurements.  The results for SMBHs are discussed in Section~\ref{sec:smbh}, using two case studies to highlight important systematic issues that can affect spin measurements.  Stellar-mass black holes in BH-XRBs are discussed in Section~\ref{sec:xrb}.  The dramatic progress that has been made recently in characterizing relativistic reverberation is summarized in Section~\ref{sec:reverberation}.   The casual reader who is interested more in the current status of spin results rather than details of the methodology may wish to skip Sections~\ref{sec:nuts} and \ref{sec:agn_cases}.   Unless otherwise stated, all spins quoted in this review are given with their 90\% error ranges.  

\section{Assumptions and Methodology of Relativistic X-ray Reflection Spectroscopy}
\label{sec:methodology}

\subsection{Basic assumptions and geometry}

In both spectroscopic methods for determining black hole spin (CF and relativistic reflection), the central assumption is that the accretion disk remains geometrically-thin, optically-thick, and radiatively efficient down to the ISCO.  As the accretion flow crosses over the ISCO, the inward radial velocity rapidly increases, the flow quickly becomes super-sonic and super-Alfvenic (hence ballistic) and, due to conservation of mass flux, the density of the flow rapidly drops.  This basic behavior is confirmed by magnetohydrodynamic simulations of thin accretion disks (Reynolds \& Fabian 2008; Penna et al. 2010).  As discussed below, the result is that the ISCO serves as the ``inner-edge" of the various observables, including the X-ray reflection features that are our focus here.   The ability to measure spin then follows directly from the fact that the location of the ISCO (in units of gravitational radii $r_g\equiv GM/c^2$) is a simple function of spin (Bardeen, Press \& Teukolsky 1972).

This central assumption is likely to be valid for sources accreting at modest rates, with bolometric luminosities in the range $10^{-2}-0.3L_{\rm Edd}$ (where $L_{\rm Edd}$ is the usual Eddington luminosity).  This includes the thermal dominant (a.k.a. high/soft) state the of BH-XRBs, as well as the more luminous broad-optical-line Seyfert galaxies and radio-galaxies.  Thermal emission from the accretion disk is clearly visible in these classes of sources, in the soft X-ray band for BH-XRBs and the optical/UV/EUV bands for AGN.  Generically, however, the spectra of these sources also show a hard X-ray power-law tail which must originate from some structure external to the accretion disk, either a disk corona or the base of a jet.

\begin{figure*}
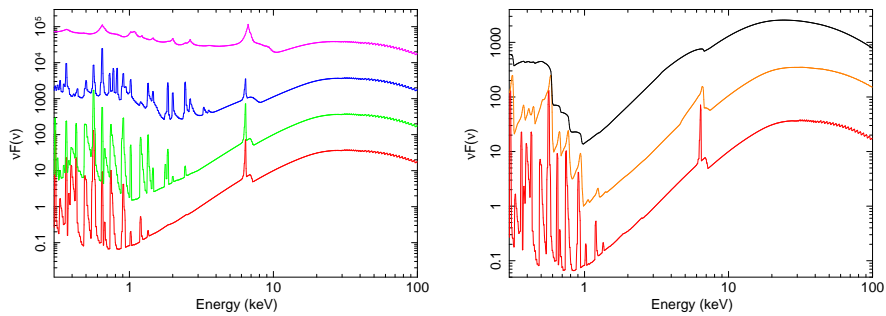

\hbox{
  \includegraphics[width=0.33\textwidth,angle=270]{reflion_color.ps}
  \includegraphics[width=0.33\textwidth,angle=270]{blur_color.ps}
  }
\caption{{\it Left panel : }Rest-frame X-ray reflection spectrum for photospheric ionization parameters of $\log\xi=0,1,2,3$ (from bottom to top).  In all cases, the irradiating source has a power-law spectrum with photon index $\Gamma=2$.  {\it Right panel : }Demonstration of the effects of relativistic Doppler/gravitational blurring on the reflection spectrum.  Curves show the $\log\xi=0$ rest-frame reflection spectrum (bottom), relativistic blurring with $r_{\rm in}=10r_g$ and emissivity index $\beta=3$ (middle curve), and extreme blurring with $r_{\rm in}=2r_g$ and $\beta=4$.  In all cases, a disk inclination of $i=30$\,degrees is assumed.}
\label{fig:reflection}    
\end{figure*}

This geometry is key to the reflection method for spin determination.   The X-ray source irradiates the underlying accretion disk with a hard X-ray continuum.  Outside of the ISCO, where the accretion disk is optically-thick and still contains combined metal ions, the irradiated surface emits a complex spectrum consisting of atomic emission features sitting on a Compton-backscattered continuum (Fig.~\ref{fig:reflection}, left).  This is the {\it X-ray reflection spectrum}, and provided that the disk photosphere is not too highly photoionized, the most prominent feature is the iron-K$\alpha$ emission line (rest-frame energy $6.4-6.97$\,keV depending upon ionization state).  Doppler shifts and gravitational redshifts, that become progressively stronger as one considers radii in the disk closer to the black hole, broaden and skew the observed features in a well-defined manner (Fig.~\ref{fig:reflection}, right).      

Within the ISCO, the density of the accretion flow plummets and it will be fully ionized by any significant X-ray irradiation of the flow.   Hence, the region within the ISCO will not contribute to the observed atomic signatures (although there may still be a Compton-reflected continuum component arising from within the ISCO).  In other words, the ISCO defines the inner edge to the region producing ``spectrally structured" X-ray reflection.  

\subsection{Operationally nuts and bolts of relativistic reflection spin determinations}
\label{sec:nuts}

Given these basics, how do we proceed at an operational level to analyze a real spectrum and extract black hole spin?  To isolate the signatures of relativistic reflection needed for the spin measurement, we must understand the spectrum of the primary continuum, the effects of absorption, and the presence of other reprocessing signatures (e.g., emission from an optically-thin photoionized wind, or X-ray reflection from distant structures such as the companion star in a BH-XRB or the molecular torus in an AGN).  

Here, I shall focus on the methodology developed for (non-blazar, Compton-thin) AGN, mentioning a few operational differences with BH-XRBs towards the end of the Section.  In a disk-dominated AGN, the primary X-ray continuum in the crucial 2--50\,keV range is very well approximated by a power-law.  This is readily understood in terms of a separation of energy-scales.  The X-rays are believed to be produced by the thermal Comptonization of soft optical/UV photons ($h\nu\sim 1-10$\,eV) by an electron population that is hot ($kT\sim 100$\,keV).  For photon energies $E$ satisfying $h\nu\ll E\ll kT$, Comptonization generically produces a power-law.    The most common and obvious complexities observed in AGN spectra are (i) a narrow iron-K$\alpha$ emission line at 6.4\,keV due to reflection/fluorescence by distant and low-velocity material, and (ii) absorption by cold and photoionized gas along the line-of-sight to the central X-ray source.    Thus, to our power-law spectral model, we add a model of cold, distant X-ray reflection (e.g., using the {\tt pexmon} model which self-consistently includes the iron-K$\alpha$, iron-K$\beta$, and nickel-K$\alpha$ lines as well as the Compton reflected continuum; Nandra et al. 2007).  To treat absorption, we add successive absorption components (each with its own ionization parameter and column density; computing using a photoionization code such as XSTAR; Kallman \& Bautista 2001) until no further improvement in the fit to the data is found.   The final step before dealing with the relativistic reflection is to add a phenomenological component (a blackbody or additional Comptonization  component) to treat the soft excess that many AGN display.

In approximately 30--50\% of type-1 AGN, applying this procedure to data with sufficient signal-to-noise will leave unmodelled residuals indicative of a broad/redshifted iron line, excess Compton reflection, i.e. a hump at $\sim 20$\,keV  (Nandra et al. 2007).  These are the signatures of relativistic reflection that we seek to characterize although, as discussed below, there are alternative explanations for the soft excess that can lead to important ambiguities.   Thus, we add to the spectral model a component describing an ionized reflection spectrum (the current industry standard is {\tt reflionx}; Ross \& Fabian 2005) that has been convolved by a ``disk-line transfer function" encoding the Doppler/gravitational redshifts from the accretion disk.  The available and commonly employed disk-line convolution models (e.g., {\tt relconv}; Dauser et al. 2010) are based on GR ray-tracing calculations within the Kerr metric, assuming a razor thin accretion disk lying in the equator of the rotating black hole.  

In addition to the black hole spin $a$ and the disk-inclination $i$, the disk-line transfer function depends upon the intensity of the irradiating flux as a function of radius, $f_X(r)$, that, of course, depends upon the nature of the primary X-ray source.   It is sometimes said that the lack of knowledge of the location and structure of the X-ray source is a major limitation for studies of relativistic reflection.  In fact, from the point of view of spin measurements, all we require is a parameterized form of $f_X(r)$ that can then be constrained by the data themselves.    Empirically, it is found that a broken-powerlaw is a good description of this irradiation profile, giving up to five additional fit parameters, the inner/break/outer radii $(r_{\rm in}, r_{\rm br}, r_{\rm out})$ and the two power-law indices $(\beta_1, \beta_2)$ defined such that in the relevant patch we have $f(X)\propto r^{-\beta}$.   As detailed above, we identify $r_{\rm in}$ with the ISCO.  Furthermore, provided that $\beta_2>2$, the spectrum will be insensitive to a sufficiently large $r_{\rm out}$. Thus, the total set of parameters describing the relativistic reflection is $\lbrace a, i, \xi, \beta_1, \beta_2, r_{\rm br}\rbrace$.    The final spectral model describing the continuum source, the distant reprocessing, absorption components, and relativistic reflection from the accretion disk, is then fitted to the data (with all interesting spectral parameters, including those describing the continuum form and the absorption components allowed to be free) and constraints on the spin can be obtained.     

There are two aspects of this methodology that must be underscored.   Firstly, in many situations, the error bar on the derived spin will be driven as much by uncertainties in the continuum/absorber parameters as by uncertainties in the form of the relativistic reflection itself.  Thus, it is crucial to allow all relevant spectral parameters to vary in the final spectral analysis, including those describing the continuum and absorption.  Secondly, when attempting to measure black hole spin, one must model the full ionized reflection spectrum.   Phenomenological models in which isolated broad iron lines are added to the primary continuum are useful for initial explorations of spectra, but are inadequate for characterizing the subtle spectral signatures of spin.

A very similar procedure is adopted in the case of BH-XRBs, although there are a few differences compared with the case for AGN driven by the fact that XRB accretion disks are hotter, the astrophysical environments are cleaner and, typically, the signal-to-noise (s/n) in the datasets is much better.   Absorption is much less of a concern in studies of the disk reflection in BH-XRBs.  Strong outflows are confined to strongly thermally-dominated states (of potential concern for CF spin measurements) but are weak or absent during states that display prominent reflection (Miller et al. 2006, 2008; King et al. 2011, 2012).  Even when present, the outflows tend to have significantly higher ionization states (such that iron has a fully stripped L-shell) and, hence, absorption tends not to introduce broad-band curvature that can confuse studies of the broad iron line.  On the other hand, the thermal disk spectrum in BH-XRB is now in the X-ray band and must be modeled if one hopes to robustly probe the broad iron line and other disk reflection signatures.   Furthermore, the disk reflection features (and especially the iron-K lines) have significant broadening even in the rest frame from Compton scattering in the hot disk atmosphere.  While this additional broadening is captured by the ionized reflection models (see top model in left panel of Fig.~\ref{fig:reflection}), there is some uncertainty in the degree of Comptonization that can affect spin determinations (see discussion of XTE~J1550--564 in Section~\ref{sec:xrb}).   Of course, the s/n is often extremely high for BH-XRBs.  Indeed, one must start to consider ``bright source" instrumental issues such as errors in the calibration of the effective area and photon pile-up.   Miller et al. (2010) has explicitly shown that pile-up tends to make a broad line narrower and, hence, always acts to decrease the inferred spin.  We return to this point in the discussion of Cygnus~X-1 in Section~\ref{sec:xrb}.

We end this Section on methodology by highlighting the complementarities and strengths of the relativistic reflection method in comparison to the CF method.   As discussed by McClintock \& Narayan (this volume), the CF method, in its essence, uses the Stefan-Boltzamann law to measure the emitting area of the accretion disk (employing a Novikov \& Thorne [1974] emission profile) and hence the size (in physical units) of the ISCO.   To apply this technique requires accurate optical/IR measurements of mass (to convert the ISCO into units of gravitational radii, $r_g=GM/c^2$), distance (in order to derive luminosities), and inclination, as well as a trustworthy absolute X-ray flux calibration.  By contrast, the relativistic transfer function underlying the reflection method is completely independent of mass, i.e., all radii are naturally scaled in units $r_g$, and one never needs to know the explicit size of $r_g$.  Hence, mass uncertainties are irrelevant to reflection-based spins.   In addition, since we are interested purely in spectral shape, the distance to the source and the absolute flux is also irrelevant (although, for AGN, the cosmological redshift is important to know).  Finally, while the inclination of the inner disk is an important quantity, it can be directly measured along with spin from the X-ray spectrum.  

\section{Supermassive black holes}
\label{sec:smbh}

\subsection{Early History}

The first quantitative attempt to constrain black spin was based on modeling of the broad iron line profile in the Seyfert galaxy MCG--6-30-15.   A long {\it ASCA} observation of this source had caught its transition to the so-called ``Deep Minimum State" in which Iwasawa et al. (1996) noticed that the iron line was too broadened and redshifted to be explained by a non-rotating black hole.  Subsequent modeling of this line profile by Dabrowski et al. (1997) concluded that $a>0.95$ under the assumption that the irradiation profile of the reflection follows a Novikov-Thorne profile.  In its original form, however, this analysis was not completely robust --- Reynolds \& Begelman (1997) showed that, in a lamp-post geometry\footnote{In the lamp-post geometry, the X-ray source is situated at some height above the disk plane on the spin-axis of the black hole.}, the combination of light-bending effects and a small penetration of the line-emitting region inside of the ISCO could render the Deep Minimum line profile consistent with Schwarzschild ($a=0$) geometry.  Young, Ross \& Fabian (1998) showed that the need for rapid spin in MCG--6-30-15 was restored when one considered the full reflection continuum, showing that the Reynolds \& Begelman [1997] scenario predicted a strong, and unobserved, absorption edge from ionized iron.  This was the first realization that modeling the full reflection spectrum, not just the iron line, was crucial for spin determinations.

The first spin-analysis to follow the modern methodology (Section~\ref{sec:methodology}) was by Brenneman \& Reynolds (2006), also for MCG--6-30-15.  Using a long and high s/n {\it XMM-Newton} observation, they confirmed the rapid spin in MCG--6-30-15, obtaining a formal limit of $a>0.98$ (although finite disk-thickness effects can relax this limit to $a>0.92$; Reynolds \& Fabian [2008]).   Wider application of these techniques to other AGN had to await the availability of sufficiently deep observations.  Over the past few years, the number of useful datasets has expanded, leading to an explosion in spin studies.  

\begin{figure*}
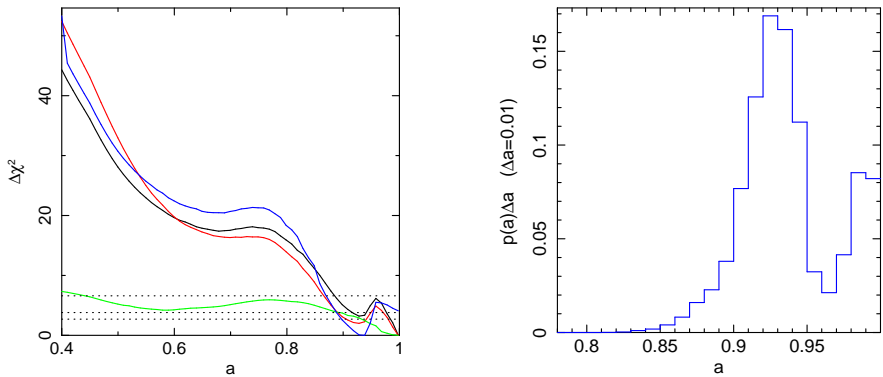

\hbox{
  \includegraphics[width=0.4\textwidth,angle=270]{ngc3783_chi2spin.ps}
  \hspace{1cm}
  \includegraphics[width=0.4\textwidth,angle=270]{ngc3783_spinprob.ps}
  }
\caption{Spin constraints on the SMBH in NGC3783 from the 2009 {\it Suzaku} observation.   {\it Left panel : }Goodness of fit relative to the best fit, $\Delta\chi^2$, as a function of the spin parameter $a$.  Different lines show the effects of different data analysis assumptions; a fiducial analysis (black), an analysis in which the warm absorber parameters are frozen at their best values (red), an analysis in which the XIS and PIN instrumental cross-normalizations are allowed to float (blue), and an analysis that ignores all data below 3\,keV.   From Brenneman et al. (2011).    {\it Right panel : }Probability distribution for $a$ as derived from a Monte Carlo Markov Chain (MCMC) analysis using the fiducial spectral model.  From Reynolds et al. (2012).}
\label{fig:ngc3783_spin}    
\end{figure*}

\subsection{Two illustrative Case Studies}
\label{sec:agn_cases}

We now discuss two case studies that demonstrate the quality, as well as possible ambiguities, of reflection-based SMBH spin measurements.  We will focus on results from one particular program, the {\it Suzaku AGN Spin Survey} (SASS), a Cycle 4--6 {\it Suzaku} Key Program (PI: Reynolds).

\begin{figure*}
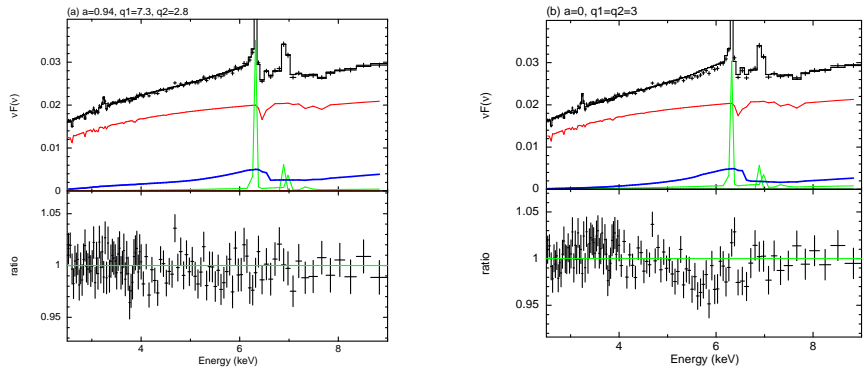

\hbox{
  \includegraphics[width=0.4\textwidth,angle=270]{ngc3783_bestfit.ps}
  \hspace{1cm}
  \includegraphics[width=0.4\textwidth,angle=270]{ngc3783_schfit.ps}
  }
\caption{Demonstration of the spectral signatures which, in practice, drive spin constraints using the {\it Suzaku} data for NGC~3783.  {\it Left panel : }Unfolded XIS data overlaid with the best fitting model (top) and the associated data/model ratio (bottom).  {\it Right panel : }Same, except that the spin parameter has been frozen at $a=0$ and (for physical consistency) the irradiation indices have been frozen at $\beta_1=\beta_2=3$.  All other parameters (including those associated with the warm absorbers) have been allowed to fit freely.   For both panels, the model components are colored as follows: absorbed power-law continuum (red), distant reflection (green), and relativistically smeared disk reflection (blue).      Figure from Reynolds et al. (2012).}
\label{fig:ngc3783_res}    
\end{figure*}

A posterchild for SASS is the bright Seyfert 1.5 galaxy NGC~3783 ($z=0.0097$).  Despite the complex (multi-zone) photoionized absorber and the presence of significant distant reflection, the relativistic reflection signatures from the inner accretion disk could be isolated and the spin constrained to be $a>0.88$ (Fig.~\ref{fig:ngc3783_spin}; Brenneman et al. 2011; Reis et al. 2012; Reynolds et al. 2012).  Figure~\ref{fig:ngc3783_res} shows the residuals when comparing the data with the best fitting spectral model ($a=0.94$) and a model in which we have forced a non-spinning black hole (and, for physical consistency, $\beta_1=\beta_2=3$) but allowed all other spectral parameters to fit.   We see that the non-spinning model is over-predicting the spectrum in the 5--6\,keV band.   There is a clear way to understand this.    In this spectrum, which includes hard-band data (up to 45\,keV), the overall fractional amount of reflection is well determined.  The question then arises whether the observed iron line is consistent with that reflection.   Consistency of the expected line strength with the data demands that the iron line is strongly broadened and hence that the black hole is rapidly spinning.  If a non-spinning black hole is imposed in the spectral fit, the iron line is ``too narrow" and, given that its strength is set by the overall amount of reflection, one over-predicts the 5--6\,keV flux.  

\begin{figure*}
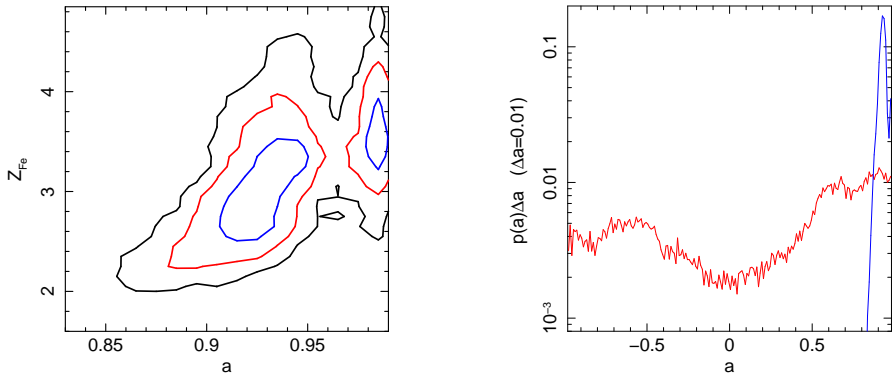

\hbox{
  \includegraphics[width=0.4\textwidth,angle=270]{ngc3783_fespin.ps}
  \hspace{1cm}
  \includegraphics[width=0.4\textwidth,angle=270]{ngc3783_fespinprob.ps}
  }
\caption{Influence of iron abundance on the measured spin, illustrated using the 2009 {\it Suzaku} data for NGC~3783 .   {\it Left panel : }Two-dimensional probability distribution for iron abundance $Z_{\rm Fe}$ and spin $a$ showing the existence of a statistical correlation between these two variables.   {\it Right panel : }Probability distribution for spin $a$ assuming a free-fitting iron abundance (blue) and an iron abundance fixed to solar values.  Figures from Reynolds et al. (2012). }
\label{fig:ngc3783_fe}    
\end{figure*}

The strength of the iron line also depends upon the iron abundance.   Thus, from the same considerations described in the previous paragraph, we would expect the spectral fit to show a statistically correlation between the spin and the iron abundance.  Our analysis of NGC~3783 finds exactly such a correlation (Fig.~\ref{fig:ngc3783_fe}, left).   In a free-fit, the spectral analysis of NGC~3783 clearly demands iron abundances that are 2--4$\times$ the cosmic average, in line with the metallicity enhancement often inferred within Seyfert nuclei on the basis of the optical/UV line ratios (e.g., Warner et al. 2004; Nagao et al. 2006).  However, if we simply impose solar iron abundances, the spectral model is a poorer fit to the data ($\Delta\chi^2=36$) and we essentially lose the ability to diagnose spin (Fig.~\ref{fig:ngc3783_fe}, right).  This exercise clearly demonstrates the need to permit non-solar iron abundance in the spectral fit if one is interested in measuring spin.   We also note that, as discussed in detail in Reynolds et al. (2012), robust spin measurements in absorbed AGN such as NGC~3783 absolutely require tabulated absorption models with adequate resolution in the $\xi$-dimension.  

While NGC~3783 is an example of an object in which we can obtain a clean and robust spin result, some objects can display spectral complexity that conspires with the medium-resolution of today's spectrometers in order to produce ambiguities in the measured spin.  These ambiguities can be particularly severe if hard X-ray data is unavailable.  We illustrate this with another of the SASS objects, the luminous Seyfert galaxy Fairall~9 ($z=0.047$).  This is an example of a ``bare" Seyfert nucleus in the sense that every pointed X-ray observation shows it to be free of intrinsic absorption, either neutral or ionized (although {\it RXTE} monitoring does show unusual X-ray ``dips" that may correspond to transient Compton-thick absorption events; Lohfink et al. 2012a).   Analysis of earlier {\it Suzaku} data by Schmoll et al. (2009) found an obvious broad iron line with a red-wing extending down to 5\,keV or possibly lower and, by imposing some restrictions on the irradiation profile, was able to estimated the black hole spin to be $a=0.6\pm 0.1$.   

\begin{figure}
\hbox{
  \includegraphics[width=0.50\textwidth]{l12_fig3.eps}
  \hspace{1cm}
  \includegraphics[width=0.3\textwidth]{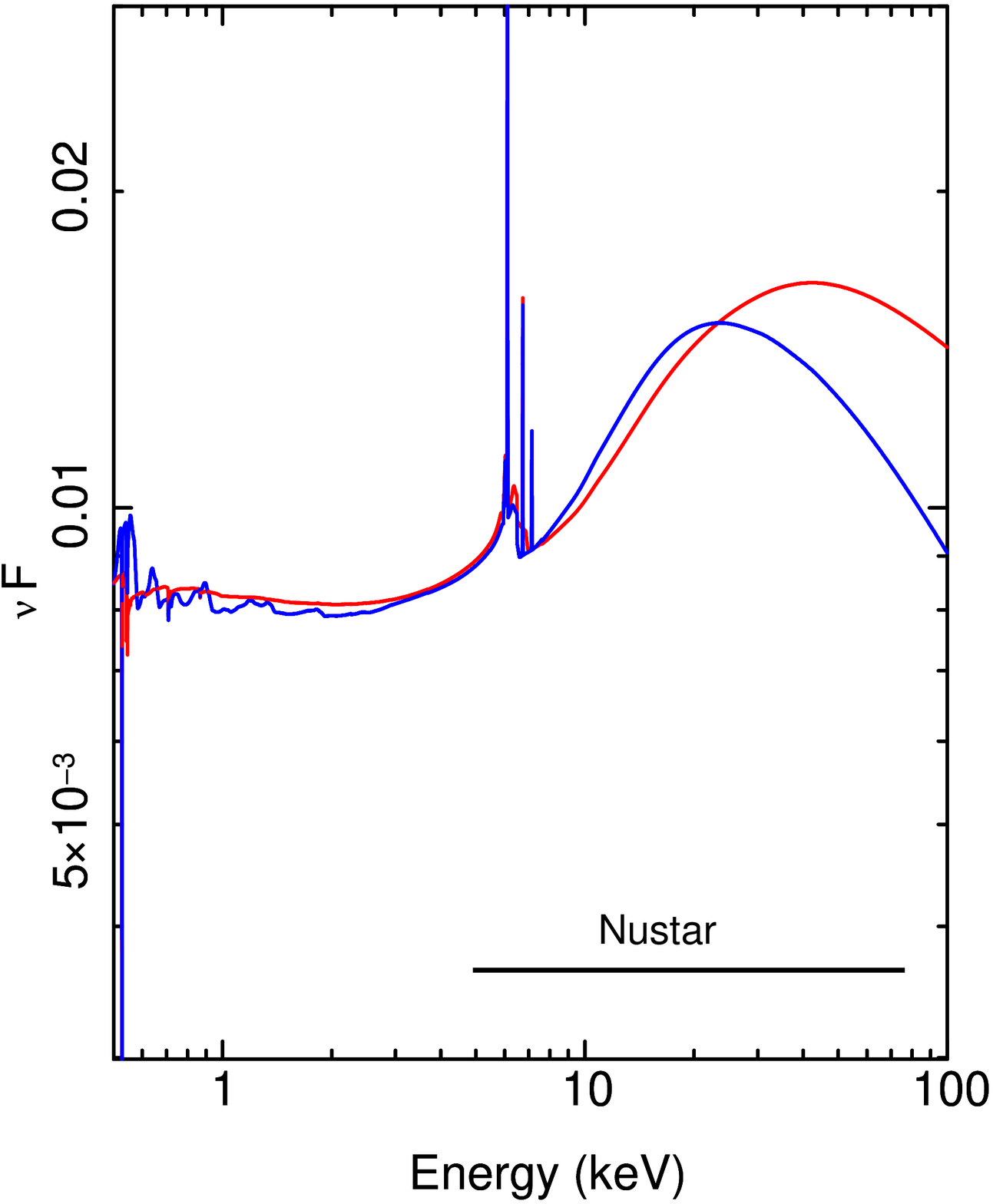}
}
\caption{{\it Left panel : }Unfolded spectra for the two {\it XMM-Newton} (blue and magneta) and two {\it Suzaku} pointings (red and black) of Fairall~9.  The 1.5-2.5\,keV band have been excluded from the {\it Suzaku} spectra due to the presence of known calibration artifacts.   Note the presence of the soft excess that appears to become more prominent as the source brightens. {\it Right panel : }Extrapolation of the two spectral models for Fairall~9 to higher energies; the red line shows the model with a thermal Comptonization soft excess, and the blue shows the case where the soft excess is described by an additional highly ionized relativistic reflection component.  Figures from Lohfink et al. (2012b).}
\label{fig:fairall9}    
\end{figure}

However, the broad line in Fairall~9 is weaker than in NGC~3783 or MCG--6-30-15 and, while it lacks absorption, there are some additional spectral complications.  Due to the higher inclination of the accretion disk, the blue-peak of the broad line is blended with narrow Fe25/Fe26 emission lines from circumnuclear photoionized material.  There is also a smooth and variable soft excess below 2\,keV (Fig.~\ref{fig:fairall9}, left).  To deal with such complexity, Lohfink et al. (2012b) accumulated all available {\it XMM-Newton} and {\it Suzaku} data (taken over the course of several years) and applied the technique of multi-epoch fitting in which physical arguments are used to tie certain spectral parameters across epochs, but all other spectral parameters are allowed to fit freely to the spectrum from each epoch.   Specifically, they tied the black hole spin, accretion disk inclination, iron abundance, and the strength of the narrow iron line components (which originate from parsec scales and hence will be stable between different observations).  This procedure finds two acceptable solutions.   In the first solution, the soft excess is described by an additional continuum component (modeled as thermal Comptonization from a moderately hot $kT\sim 20$\,keV plasma) and the relativistic disk reflection, driven by the modestly broad iron line, implies a modestly spinning black hole ($a=0.52^{+0.19}_{-0.15}$, completely consistent with the previous analysis of Schmoll et al. 2009).   Interestingly, such a solution demands a very high iron abundance ($Z_{\rm Fe}\sim 10Z_\odot$) which would drive one to consider somewhat exotic phenomena such as radiative-levitation to enhance the photospheric iron content of the inner disk (Reynolds et al. 2012).    However, a second solution exists which is a slightly worse description of the data but still statistically acceptable overall.  Here, the soft excess is described by a highly-ionized ($\xi\sim 10^3$) relativistic reflection component from the inner accretion disk.  In order to model the modest breadth of the broad iron line, the same disk must have low ionization patches that become more dominant at larger radii.  This solution has an iron abundance that is slightly sub-solar, but the need to produce a very smooth soft excess from a structured reflection spectrum drives the fit to a rapidly-spinning black hole ($a>0.93$).

These two solutions predict very different behavior above 20\,keV (Fig.~\ref{fig:fairall9}, right).  In principle, the PIN instrument on {\it Suzaku} should have distinguished these models.  Unfortunately, due to pointing and aspect problems during the observation (discussed in Lohfink et al. 2012b), the PIN data cannot be normalized with respect to the XIS data and, together with the low s/n of Fairall~9 in the PIN detector, provides no additional constraints.  This experience underscores the importance of data above 10\,keV even when one is attempting to extract spin from signatures below 10\,keV.     We look forward to {\it NuSTAR} and {\it Astro-H} studies of Fairall~9 to squash these ambiguities.

\subsection{Summary of Current SMBH Spin Measurements}
\label{sec:smbh_current}

\begin{table}
\caption{Summary of published AGN/SMBH spin measurements that pass the quality-criteria laid out in Section~\ref{sec:smbh_current}.  All measurements are based upon {\it XMM-Newton} and/or {\it Suzaku} data.  Reflecting the conventions in the primary literature, all masses are quoted with $1\sigma$ error bars whereas spins are quoted with $90\%$ error ranges.   Column (1) lists the common name for the AGN, Columns (2) and (3) give the mass and spin, respectively, and Column (4) gives the references for these mass/spin measurements.  Key to references: Be11=Bennert et al. (2011); BR06=Brenneman \& Reynolds (2006); Br11=Brenneman et al. (2011); Fa13=Fabian et al. (2013); Ga11=Gallo et al. (2011); Go12=Gonz\'ales-Martin \& Vaughan (2012); Lo12=Lohfink et al. (2012b); Ma08=Malizia et al. (2008); Mc05=McHardy et al. (2005); Mi09=Miniutti et al. (2009); Pe04=Peterson et al. (2004); Pa12=Patrick et al. (2012); Ta12=Tan et al. (2012); Wa13=Walton et al. (2013); Zo10=Zoghbi et al. (2010); ZW05=Zhou \& Wang (2005).}
\label{tab:1}       % Give a unique label
\begin{center}
\begin{tabular}{lcccl}
\hline\noalign{\smallskip}
Object  & Mass ($\times 10^6M_\odot$) & Spin & Mass/Spin References  \\
\noalign{\smallskip}\hline\noalign{\smallskip}
Mrk335  		& $14.2\pm 3.7$	& $0.83^{+0.09}_{-0.13}$	& Pe04/Wa13 \\
IRAS~00521--7054 & ---			& $>0.84$				& --/Ta12\\
Tons180 		& $\sim 8.1$		& $0.92^{+0.03}_{-0.11}$	& ZW05/Wa13 \\
Fairall~9		& $255\pm 56$		& $0.52^{+0.19}_{-0.15}$	& Pe04/Lo12 \\
Mrk359		& $\sim 1.1$ 		& $0.66^{+0.30}_{-0.54}$	& ZW05/Wa13 \\
Mrk1018		& $\sim 140$		& $0.58^{+0.36}_{-0.74}$	& Be11/Wa13 \\
1H0419-577 	& $\sim 340$		& $>0.89$				& ZW05/Wa13\\
Ark120		& $150\pm 19$		& $0.64^{+0.19}_{-0.11}$	& Pe04/Wa13\\
Swift~J0501.9-3239 & ---			& $>0.99$				& --/Wa13 \\
1H0707-495	& $\sim 2.3$		& $>0.97$				& ZW05/Zo10\\
Mrk79		& $52.4\pm 14.4$	& $0.7\pm 0.1$		& Pe04/Ga11\\
Mrk110		& $25.1\pm 6.1$	& $>0.89$				& Pe04/Wa13\\
NGC3783 	& $29.8\pm 5.4$ 	& $>0.88^*$ 			& Pe04/Br11 \\
NGC4051		& $1.91\pm0.78$	& $>0.99$				& Pe04/Pa12 \\
RBS1124		& ---				& $>0.97$				& --/Wa13\\
IRAS13224--3809 & $\sim 6.3$	& $>0.987$			& Go12/Fa13\\
MCG--6-30-15	& $2.9^{+1.8}_{-1.6}$& $a>0.98$			& Mc05/BR06\\
Mrk841		& $\sim 79$		& $>0.52$				& ZW05/Wa13\\
Swift~J2127.4+5654 & $\sim 1.5$	& $0.6\pm 0.2$			& Ma08/Mi09\\
Ark564 		& $\sim 1.1$		& $0.96^{+0.01}_{-0.11}$	& ZW05/Wa13 \\
\noalign{\smallskip}\hline
\end{tabular}
\end{center}
$^*$Note that Patrick et al. (2012) find a spin of $a<-0.35$, clearly discrepant with the results of Brenneman et al. (2011) report here.  However, this object has a high column density warm absorber.  The modeling of this absorber, specifically the resolution of the table-models used to characterize the absorber, appears to be the root cause of the spin discrepancy.  
\label{tab:spinsummary}
\end{table}

The past couple of years has seen an explosion in the number of published SMBH spin measurements based on relativistic reflection spectra, including two major works that use the public data archives to compile samples of SMBH spins (Patrick et al. 2012; Walton et al. 2013).   Drawing on the published refereed literature at the time of writing (January 2013), Table~\ref{tab:spinsummary} compiles the 20 SMBH/AGN spin measurements that satisfy the following ``quality control" criteria:
\begin{enumerate}
\item Spin measurement are based on fits of the fully ionized reflection spectrum (not just an isolated broad iron line).  In practice, all of these measurements have employed the Ross \& Fabian (2005) model {\tt reflionx}.
\item The iron abundance characterizing the disk reflection is allowed to be a free parameter in the spectral fit.  This is important given the abundance-spin partial degeneracy noted in Section~\ref{sec:agn_cases}.
\item The inclination of the accretion disk is allowed to be a free parameter in the spectral fit and can be constrained.  In cases where the inclination cannot be constrained, the spin constraint must be driven by some aspect of the spectrum other than the broad iron line (e.g., the soft excess) and one must be concerned about lack of uniqueness of the spectral model.
\item The irradiation index $\beta$ is allowed to be a free parameter and converges on a value $\beta>2$.   In cases where the best-fit implies $\beta<2$, the X-ray reflection is dominated by the {\it outer} accretion disk and so we would not expect to be sensitive to spin.
\end{enumerate}
In addition to these quality-control criteria, we have rejected two sources from the Walton et al. (2013) sample; Mrk~509 has a {\it Suzaku}-based spin value that is not robust to the treatment of the PIN normalization, and PDS~456 has a spin solution that appears to be double-valued.   We have included Fairall~9, listing its intermediate spin value from Lohfink et al. (2012b; based on the strict implementation of the methodology discussed in Section~\ref{sec:methodology}), but we note the ambiguity of this particular spin value (see Section~\ref{sec:agn_cases}).

While this is far from a well-defined or complete sample of objects, we can already draw some interesting conclusions.   Firstly, there are a large number of objects with rapidly spinning black holes.   For the most rapid spinners ($a>0.9$), the numbers must be taken with a pinch of salt --- the reported spins are the formal results (with statistical errors) that arise from the application of our standard methodology which assumes a razor thin Keplerian accretion disk.   Using the radiation-pressure disk theory of Shakura \& Sunyaev (1973), we expect the inner disks of these Seyfert galaxies, most of which are between 1--30\% of their Eddington luminosities, to have thicknesses of $h\sim 0.1-1r_g$.   Hence, especially for sources with rather large Eddington ratios, we may expect finite disk thickness effects at the ISCO to become relevant when the ISCO becomes small, i.e. for large spins.  Further work is needed to fully characterize this systematic error, but preliminary considerations by Reynolds \& Fabian (2008) suggested that finite-thickness effects could relax the limits on the ``true" spin value in these extreme cases to $a>0.9$.

\begin{figure}
\centerline{
  \includegraphics[width=0.80\textwidth]{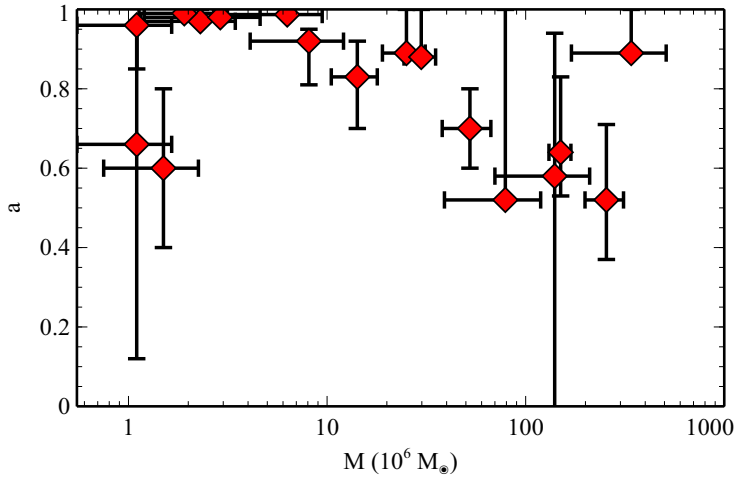}
}
\caption{Plot of the SMBH mass $M$ and spin $a$ from the sample listed in Table~\ref{tab:spinsummary}.   Reflecting the conventions in the primary literature, all masses are marked with $1\sigma$ error bars whereas spins are marked with $90\%$ error ranges.   When no error estimate is available for the mass, we have assumed an error of $\pm 0.5M$.  }
\label{fig:smbh_spins}    
\end{figure}

Still, it seems clear that a significant number of SMBHs in these AGN are rapidly spinning ($a>0.8$), at least in the mass range $10^6-{\rm few}\times 10^7M_\odot$.  This suggests that the last mass doubling of these SMBHs occurred via coherent accretion and not chaotic accretion (King \& Pringle 2006) or SMBH-SMBH major merger (Volonteri et al. 2005).  Interestingly, as shown in Fig.~\ref{fig:smbh_spins} there are hints that the most massive black holes in this sample ($>10^8M_\odot$) as well as the least massive black holes ($M<2\times 10^6m_\odot$) may have more modest spins.   If these trends hold up in more rigorous analyses, they would provide direct evidence for the increased role of chaotic accretion and/or major mergers at these two extreme ends of the SMBH mass spectrum.

The second noteworthy point is that all of the AGN in Table~\ref{tab:spinsummary} are radio-quiet, and many of them are hosted by late-type (i.e., spiral disk) galaxies.  This appears to conclusively reject the notion that black hole spin drives the radio-quiet/radio-loud dichotomy and its association with host galaxy type (Sikora, Stawarz \& Lasota 2007).   

\section{Stellar mass black holes}
\label{sec:xrb}

\subsection{Early History}

The first black hole broad iron line was discovered by Barr, White \& Page (1985) in the {\it EXOSAT} spectrum of the BH-XRB Cygnus X-1.    These authors interpreted the line broadening as the effects of Compton scattering of iron emission line photons in the hot disk atmosphere.    It was several years before this result was re-interpreted as broadening due to Doppler/gravitational shifts in the innermost regions of the black hole accretion disk (Fabian et al. 1989). 

Due to bright-source limitations by CCD spectrometers, relativistic reflection studies in BH-XRB lagged those of AGN during the {\it ASCA} era.  However, this situation changed with the launch of {\it Chandra}, {\it XMM-Newton} at the turn of the millennium.   The first robust sign of spin in the reflection spectrum of a BH-XRB was found in the {\it XMM-Newton}/EPIC-pn data of XTE~J1650--500 (Miller et al. 2002), and quantitative constraints on spin using an early version of the methodology described in \S\ref{sec:methodology} were derived for the BH-XRB GX339--4 (Miller et al. 2004).  

\subsection{Challenges and Rewards}

To measure the spin of a stellar mass black hole in a BH-XRBs, one must confront several challenges.   Typically, the crucial 2--10\,keV part of the X-ray spectrum contains contributions from both the thermal (blackbody) disk emission and the Comptonized hard X-ray tail.   When applying disk-reflection methods to determine spin, one must model the thermal disk emission and be cognizant of the fact that any mis-modeling of the continuum may affect the inferred spin.   Similarly, when applying the CF method, one attempts to select spectral states that are overwhelmingly dominated by the thermal disk emission but any remaining Comptonization of that emission (beyond that already included in the disk atmosphere models) will skew the spin results.  

Additional complications are caused, ironically, by the fact that these sources are so bright.  In many of today's more capable X-ray (CCD) spectrometers, special instrumental modes are required in order to observe such sources while avoiding telemetry saturation as well as photon-pileup and other more subtle instrumental effects.  Thus, unlike the case for AGN, one is often in the realm where systematic uncertainties related to both the spectral models as well as the calibration of the spectral data dominate statistical (photon) errors; there is the danger that one will end up with very precise but inaccurate spin measures.  These concerns are valid for both the reflection- and CF-methods.  

Once these challenges are overcome, however, BH-XRBs offer a tremendous reward --- the possibility of measuring spin using multiple techniques and thereby validate the underlying assumptions of all spin measurements.   For this reason, BH-XRBs have deservedly received significant attention.

\begin{table}
\caption{BH-XRBs that have reflection-based spin determinations, together with the CF fitting result when one exists.   Spins are listed with their 90\% error ranges.  Key to reference : Bl09=Blum et al. (2009); Fa12=Fabian et al. (2012); Go11=Gou et al. (2011); Hi=Hiemstra et al. (2011); Mc06=McClintock et al. (2006); Mi09=Miller et al. (2009); Rei09=Reis et al. (2009); Rei11=Reis et al. (2011); Rei12=Reis et al. (2012b); Sh06=Shafee et al. (2006); St11=Steiner et al. (2011); St12=Steiner et al. (2012).  }
\label{tab:1}       % Give a unique label
\begin{center}
\begin{tabular}{lccl}
\hline\noalign{\smallskip}
Object  		& Spin from Reflection 	& Spin from CF 	& References  \\
\noalign{\smallskip}\hline\noalign{\smallskip}
4U~1543--475 & $0.3\pm 0.1$			& $0.8\pm 0.1$	& Mi09/Sh06\\
Cygnus~X-1 	& $>0.95^*$				& $>0.95$			& Fa12/Go11\\
GX339--4 	& $0.94\pm 0.02$		& ---				& Mi09/--\\
GRS1915$+$105 &	$>0.97^\dagger$			& $>0.95$			& Bl09/Mc06\\
GRO~J1655--40 & $>0.9^*$		& $0.7\pm 0.1$ 	& Rei09/Sh06\\
LMC X~1		& $>0.55$				& $>0.87$			& St12/Go09\\
MAXI~J1836--194 & $0.88\pm 0.03$	& ---				& Rei12/-- \\
SAX~J1711.6--3808 & $0.6^{+0.2}_{-0.4}$ & ---			& Mi09/--\\
Swift~J1753.5--0127 & $0.76^{+0.11}_{-0.15}$ & ---			& Rei09/--\\
XTE~J1550--564 & $0.33-0.77^\ddagger$ & $0.34^{+0.37}_{-0.45}$ & Mi09/St11\\
XTE~J1650--500 & $0.79\pm 0.01$		& ---				& Mi09/--\\
XTE~J1652--453 & $0.45\pm 0.02$		& ---				& Hi11/-- \\
XTE~J1752--223 & $0.52\pm 0.11$		& ---				& Rei11/--\\
XTE~J1908$+$094 & $0.75\pm 0.09$	& ---				& Mi09/--\\
\noalign{\smallskip}\hline
\end{tabular}
\end{center}
$*$We note that Mi09 quote a spin value of $a<0.05$.   However, it appears likely that the {\it XMM-Newton} data on which this measurement rested were affected by photon-pileup (J.M.Miller, private communication).  See further discussion in Section~\ref{sec:xrb}\\ 
$\dagger$We quote the high-spin model of Bl09 which employs a broad-band fit of a broken power-law primary continuum and a corresponding reflection continuum modeled with a blurred {\tt pexriv} (Magdziarz \& Zdziarski 1995).  The more self-consistent {\tt reflionx} model (Ross \& Fabian 2005) could not be adequately applied to the broad-band data due to the complex continuum.  \\
$\ddagger$Result combines the work of Mi09 and St11 to develop a composite error range that encompasses systematic uncertainties of the reflection model.  See discussion in Section~\ref{sec:xrb}.
\label{tab:xrbspin}
\end{table}

\subsection{Summary of Current Stellar-Mass Black Hole Spin Measurements}

%$0.76\pm 0.01\hspace{0.1cm}(0.55^{+0.15}_{-0.22})$

Table~\ref{tab:xrbspin} draws together the BH-XRBs with published spins based on the relativistic reflection method.  Of the six objects with both reflection- and CF-based spin measurements, there is general concordance for four.   Clear discrepancies remain for two objects, 4U~1543--475 and GRO~J1655--40, the origin of which remain unclear.  

Two of the systems in which there are concordant spin measurements deserve additional discussion.  For Cygnus~X-1, there is now general agreement between reflection-based measurements (Fabian et al. 2012) and CF-measurements (Gou et al. 2011) that the black hole is a rapid spinner, $a>0.95$.  Again, this is the formal result and the finite disk thickness considerations discussed in Section~\ref{sec:smbh_current} may weaken this limit to $a>0.9$.   However, it must be noted that early spin measurements using the reflection method suggested a small spin ($a<0.05$).  It now seems apparent that these early {\it XMM-Newton} data on which this measurement was based were inflicted by photon-pileup effects (J.M.Miller, private communication) that tend to narrow line features and hence decrease inferred spins (Miller et al. 2010).  Analysis of more recent {\it XMM-Newton} data using modified timing mode to handle the high count rate found two solutions for spin (Duro et al. 2012).  Assuming an irradiation profile described by a single power-law, the Duro et al. analysis suggested either an intermediate spin with a steep irradiation profile ($\beta>4$) or a rapid spin with a standard irradiation profile ($\beta\sim 3$).   Guided by the realization that steep profiles correspond to either extreme light-bending effects or inner-disk torques, both of which require rapid spins, these authors rejected the intermediate spin solution.   Finally, Fabian et al. (2012) examined {\it Suzaku} data and showed that rapid spin is favored unambiguously if one adopts a broken power-law form for the irradiation profile, a form that is verified by a non-parametric determination of the irradiation.

XTE~J1550--564 is another interesting case.   The CF method yields a rather broad range of allowable spins, $-0.11<a<0.71$ at the 90\% confidence level (Steiner et al. 2011).  Miller et al. (2009) apply the reflection method to {\it ASCA}/GIS data for this object, using the standard {\tt reflionx} (Ross \& Fabian 2005) ionized reflection model, and find a spin slightly above the upper end of the CF range, $a=0.76\pm 0.01$.   However, motivated by the fact that, for the highly ionized disks found in BH-XRBs, the iron line can be appreciably Compton broadened, Steiner et al. (2011) re-examined this object using a very high-density, hot, variant of the reflection model ({\tt refbhb}; Ross \& Fabian 2007).  They find that the additional Compton broadening of the iron line present in {\tt refbhb} leads to a slightly lower best fit spin, $a=0.55^{+0.15}_{-0.22}$.  The fact that the spin does not change dramatically implies that the broadening of the reflection spectrum is dominated by the disk dynamics (Doppler/gravitational), with Compton-broadening having a noticeable but sub-dominant effect.   Given that the true density and vertical structure of the X-ray photosphere is still uncertain, we take these two reflection results as bracketing the possible systematic error introduced by the reflection model, giving us the range of $a=0.33-0.77$ that we quote in Table~\ref{tab:xrbspin}.

We end this section with a brief discussion of the astrophysical implications.  Given that it is difficult to conceive of an astrophysical scenario whereby a stellar-mass black hole in a binary system can accrete a substantial fraction of its mass, we conclude that the measured spin must be close to the birth spin of the black hole.    Furthermore, irrespective of whether we use the reflection- or CF-based spin measures, all of the stellar mass black holes examined to date have spin parameters that greatly exceed estimates for the natal spin parameters of neutron stars ($a_{\rm NS}\sim 0.01-0.03$, Miller, Miller \& Reynolds 2011).  This constitutes direct evidence for differences in the core-collapse supernovae that have produced the black hole and neutron star populations.   One possibility is that core collapse generically forms rapidly-spinning compact objects but, in the case of neutron stars, there is speedy reduction in the angular momentum from either gravitational wave loses due to $R$-mode instabilities (Andersson et al. 1999) or a magnetic propeller mechanism.  

\section{Beyond Spectroscopy : The Age of Relativistic Reverberation Mapping}
\label{sec:reverberation}

We conclude this chapter with a brief discussion of an exciting frontier, the recent discovery of relativistic iron line reverberation.

A key aspect of the geometry underlying the X-ray reflection picture is that the hard X-ray source is external to the optically-thick flow --- i.e., there is a finite distance between the emission site for the hard X-rays and the X-ray reflector.  It is straightforward to see that variations in the luminosity of the primary X-ray source will drive variations in the strength of the observed X-ray reflection with a time-delay related to the light travel time from the source to the disk.  The full energy-dependent transfer function that relates continuum fluctuations with changes in the broad iron line encodes a tremendous amount of information including the location of the X-ray source and the spin parameter of the black hole (Reynolds et al. 1999).  For this reason, relativistic iron line reverberation has been a focus of proposals for future high-throughput X-ray observatories such as {\it LOFT} (Feroci et al. 2012).

While measurements of the full transfer function must await future observatories, recent analyses of {\it XMM-Newton} data have already found evidence for time delays between the continuum emission and relativistic reflection.    The narrow-line Seyfert-1 galaxy 1H0707--495 has a strong soft excess that can be explained as soft reflection from an ionized relativistic disk.   This interpretation is bolstered by spectral structure in the soft excess that resembles a relativistically-broadened iron-L line (Fabian et al. 2009).   Fabian et al. (2009) discovered that, at intermediate timescales, the soft-excess lags the 2--4\,keV continuum band by $\sim 30$\,s (also see detailed follow-up work by Zoghbi et al. 2010).  Since the continuum source itself always displays a pattern of hard-lags-soft (``hard lags"), this pattern of soft-lags-hard (``soft lags") is best interpreted as reverberation delays due to a displacement of the variable X-ray source from the disk (however, see alternative interpretation of these soft-lags by L.Miller et al. 2010, and a discussion of the problems with this alternative model by Zoghbi, Uttley \& Fabian 2011).   Since the initial discovery of soft-lags in 1H0707--495, they have been found in a  number of other AGN (Zoghbi \& Fabian 2011; De~Marco et al. 2012; Fabian et al. 2012b; Kara et al. 2013).  

\begin{figure*}
\centerline{
  \includegraphics[width=0.90\textwidth]{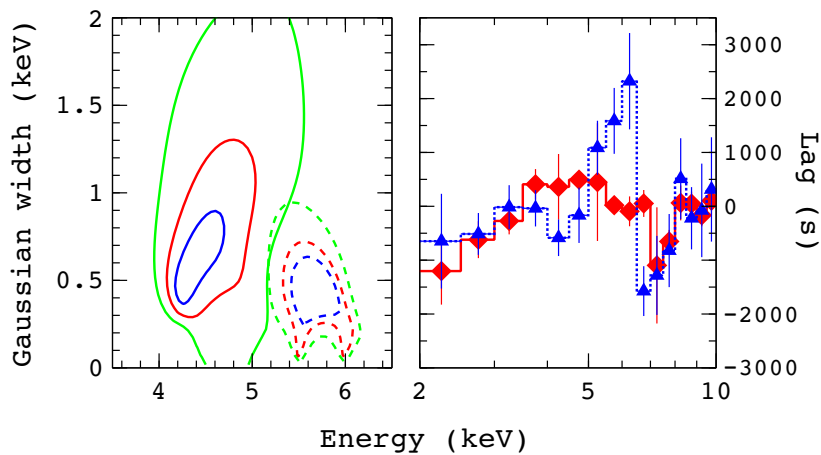}
}
\caption{Detection of lags in the broad iron line in the Seyfert galaxy NGC~4151.  {\it Right panel : }Energy lags as a function of energy for two Fourier frequency ranges, $f<2\times 10^{-5}$\,Hz (blue triangles) and $f=(5-50)\times 10^{-5}$\,Hz (red diamonds).  {\it Left panel : }Error contours from a Gaussian fit to the pag-spectra, showing the 68\%, 90\% and 95\% confidence levels.   Continuous lines refer to the $(5-50)\times 10^{-5}$\,Hz lags, and dashed lines refer to the $f<2\times 10^{-5}$\,Hz lags.  Figures from Zoghbi et al. (2012).}
\label{fig:ngc4151}    
\end{figure*}

Very recently, Zoghbi et al. (2012, 2013) has used similar techniques to find reverberation lags in the broad iron lines of NGC~4151, NGC~7314 and MCG--5-23-16 (some of the brightest Seyfert-1 galaxies in the iron-K band).  A skewed broad iron line can be clearly seen when the lags are examined as a function of energy (see NGC4151 results in Fig.~\ref{fig:ngc4151}).  Furthermore, the shape of this ``lag spectrum" is a function of Fourier frequency, with lower-frequency variations showing a narrower line with longer time-lags.    This is exactly as expected from reverberation theory, suggesting that we are gaining our first glimpse at the structure of the transfer function.  Putting this together suggests that the variable X-ray source is located above the accretion disk, possibly on the black hole spin axis, at a height of a few$\times r_g$.   This positive detection of iron line reverberation with {\it XMM-Newton} shows conclusively that future observatories, such as {\it LOFT}, will be able to  make tremendous progress in studies of strong gravity using relativistic reverberation in AGN.  

\section{Conclusion}
\label{conclusion}

X-ray reflection signatures from the inner accretion disk of BH-XRBs and AGN give us one of the most direct probe of black hole spin.  In this Chapter, I have reviewed the assumptions that underlie this technique as well as the detailed methodology as applied to both stellar-mass black hole systems and supermassive black holes.   The key points to take away from this discussion are:
\begin{enumerate}
\item Relativistic reflection spectroscopy can be applied to derive spin across the full mass range of astrophysical black holes.   A major strength of this method is that one does not need to know the black hole mass or the distance of the system in order to derive spin.  Furthermore, one does not need prior knowledge of the inclination of the inner accretion disk (it is a measured quantity along with the spin).
\item There has been an explosion of SMBH spin measurements in recent years.   Applying some objective quality-control criteria to results in the literature, we obtain a list of 19 AGN with SMBH spin measurements.  It is clear that a large fraction of SMBHs are rapidly-spinning, although there are tentative hints of a more slowly spinning population at high ($M>5\times 10^7M_\odot$) and low ($M<2\times 10^6M_\odot$) mass.
\item Stellar-mass black holes in BH-XRBs provide the invaluable opportunity to measure spin with multiple techniques, thereby allowing us to validate the basic assumptions of the various methodologies.   Comparing relativistic reflection results to those derived from continuum fitting, there is generally concordance (exceptions being 4U~1543--475 and GRO~J1655--40).  
\item The recent discovery of relativistic reverberation, especially of time delays between continuum fluctuations and the broad iron-K line, has opened a new frontier.  In addition to definitively confirming the broad iron line paradigm, this proves that future large-area X-ray observatories will make tremendous progress in studies of strong gravity using strong-field reverberation mapping. 
\end{enumerate}

\begin{acknowledgements}
I thank Andy Fabian, Anne Lohfink, Jon Miller and Abdu Zoghbi for invaluable discussion during the writing of this review. I gratefully acknowledge support from NASA under ADAP grant NNX12AE13G.
\end{acknowledgements}

%\bibliographystyle{aps-nameyear}      % basic style, author-year citations
\nocite{*}


\end{document}